\documentclass[fleqn,12pt,twoside]{article}
\usepackage[headings]{espcrc1}

\readRCS
$Id: espcrc1.tex,v 1.2 2004/02/24 11:22:11 spepping Exp $
\usepackage{graphicx}
\usepackage[figuresright]{rotating}

\newcommand{\AmS}{{\protect\the\textfont2
  A\kern-.1667em\lower.5ex\hbox{M}\kern-.125emS}}

\newcommand{\bm}[1]{\mbox{\boldmath$#1$}}

\title{Quasiparticle picture of quarks near chiral transition
       at finite temperature} 

\author{M. Kitazawa
        \address{RIKEN BNL Research Center, 
                 Brookhaven National Laboratory, Upton, NY 11973, USA},
        T. Kunihiro
        \address{Yukawa Institute for Theoretical Physics,
                 Kyoto University, Kyoto, 606-8502, Japan},
        and 
        Y. Nemoto
        \address{Department of Physics, Nagoya University,
        Nagoya, 464-8602, Japan}
        }
       
\runtitle{Quasiparticle picture of quarks at finite temperature}
\runauthor{M. Kitazawa, T. Kunihiro and Y. Nemoto}

\begin{document}

\maketitle

\begin{abstract}
We investigate, using a chiral effective model,
the quark spectrum in the critical region of the chiral transition
focusing  on the effect of the possible
mesonic excitations in the quark-gluon plasma phase.
We find that there appears a novel 
three-peak structure in the quark spectra.
We elucidate 
the mechanism of the appearance of the multi-peak
structure with the help of a Yukawa model
with an elementary boson.
\end{abstract}

\section{Introduction}

It is an intriguing problem with a rather long history to
explore the possible existence of hadronic excitations
in the quark-gluon plasma (QGP) phase\cite{HK85,MS86,Lattice}.
The existence of the hadronic (mesonic) excitations
 in the light-quark sector
was first predicted as being the soft mode associated to
the chiral transition\cite{HK85}.
The recent lattice QCD simulations also suggest the existence
of the mesonic bound states in heavy-quark sector 
in the QGP phase\cite{Lattice}.

In this work, we investigate the effect of these bosonic excitations
on the quark quasi-particle picture in the QGP phase\cite{KKN06,KKN06b}.
We first explore how the soft mode of
the chiral transition\cite{HK85} affects the quark spectrum in the QGP phase
using the Nambu-Jona-Lasinio(NJL) model,
and show that a novel three-peak structure 
is formed in the quark spectrum 
near the critical temperature $T_c$\cite{KKN06}.
We  shall elucidate  the
mechanism and show the universality of the emergence of the
three-peak structure in the fermion spectral function,
employing a Yukawa model 
composed of a massless fermion and a massive boson\cite{KKN06b}:
We show that the complicated spectra originate from the 
mixing between a quark (anti-quark) and an anti-quark hole
(quark hole) caused by a ``resonant scattering'' 
of the quasi-fermions with the thermally-excited massive boson.

\section{Quark spectrum in NJL model}

To explore the quark matter near $T_c$,
we first employ the two-flavor Nambu--Jona-Lasinio (NJL) model
in the chiral limit
\begin{eqnarray}
  \mathcal{L}=\bar{\psi} i \partial \hspace{-0.53em} / \psi
  + G_S [(\bar{\psi} \psi)^2 + (\bar{\psi}i\gamma_5\vec{\tau}\psi)^2],
\end{eqnarray}
with the coupling constant $G_S=5.5$ GeV${}^{-2}$ 
and the three dimensional cutoff $\Lambda=631$ MeV. 
This model gives a second order phase transition at $T_c=193.5$MeV
for vanishing quark chemical potential,
and the soft modes of the phase transition are 
dynamically formed in the scalar and pseudoscalar channels.
The spectral function in these channels
shows a pronounced peak in the time-like region,
the peak position and the width of which
become smaller and vanishes eventually,
as $T$ approaches $T_c$\cite{HK85,KKN06}. 

\begin{figure}
\begin{center}
\includegraphics[width=0.5\textwidth]{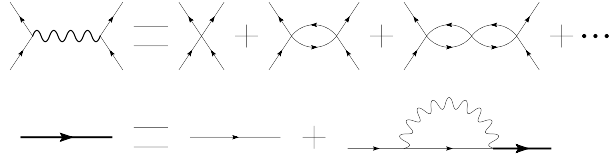}
\end{center}
\vspace*{-.5cm}
\caption[]{
The quark propagator in the RPA.
The wavy line denotes the soft modes of the chiral transition.
}
\label{fig:diagram}
\end{figure}

Now notice that 
 the soft modes composed of quarks can in turn contribute to
the quark self-energy $\Sigma$,
which may be given in the random phase approximation (RPA)
as shown diagrammatically in Fig.~\ref{fig:diagram}:
In the imaginary time formalism it reads,
\begin{eqnarray}
  \tilde{\Sigma}(\bm{p},\omega_n) =
  T\sum_{m}\int\frac{d^3 q}{(2\pi)^3} 
  \left( {\cal D}_\sigma(\bm{q},\nu_m) 
  + 3{\cal D}_\pi(\bm{q},\nu_m) \right)
  \mathcal{G}_0(\bm{p}-\bm{q},\omega_n-\nu_m ),
\end{eqnarray}
where ${\cal D}_{\sigma,\pi}(\bm{p},\nu_n)$ are
the Matsubara propagators of the scalar and pseudo-scalar channels 
and ${\cal G}_0(\bm{p},\omega_n)$
the free quark propagator.

\begin{figure}
\vspace*{-.2cm}
\begin{center}
\raisebox{6mm}{
\includegraphics[width=0.33\textwidth]{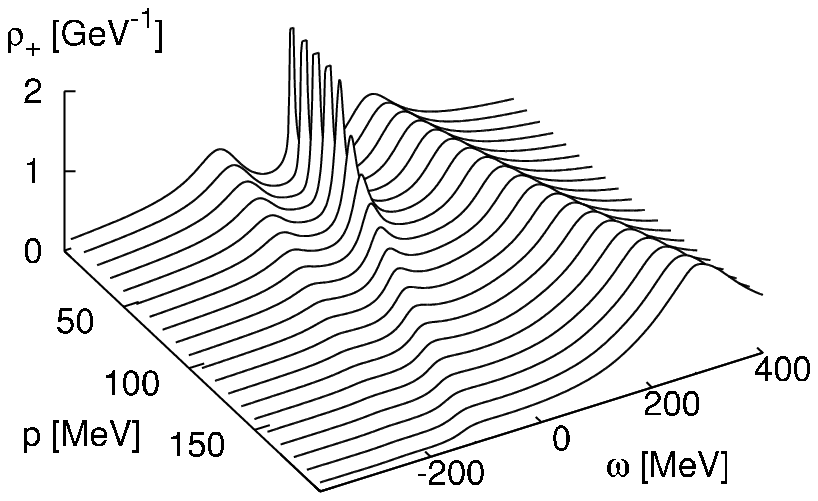}}
\includegraphics[width=0.42\textwidth]{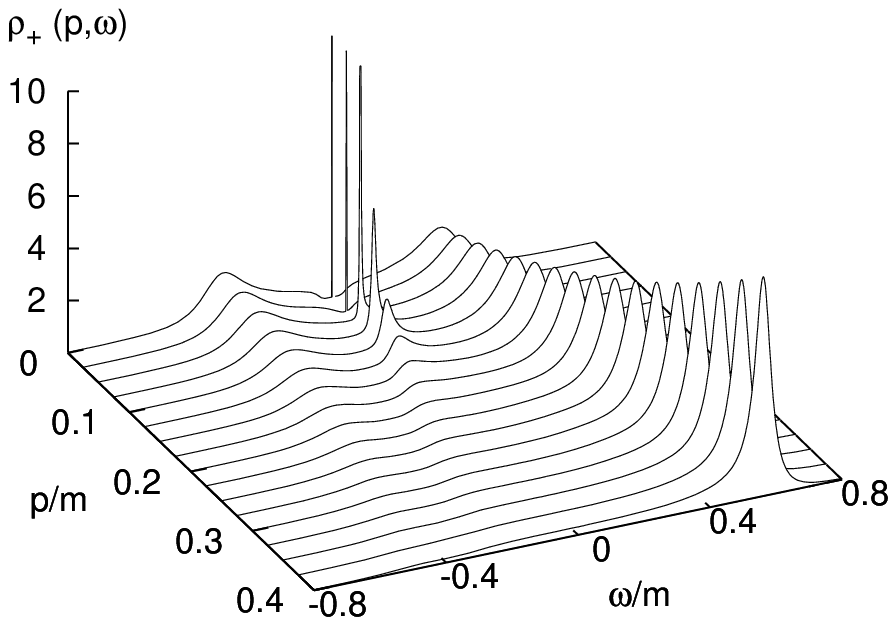}
\end{center}
\vspace*{-13mm}
\caption[]{
The quark spectral function 
$\rho_+(\bm{p},\omega)$ in the NJL and Yukawa models.
The left and right panels show $\rho_+(\bm{p},\omega)$ 
in the NJL model for 
$\varepsilon \equiv (T-T_c)/T_c = 0.1$\cite{KKN06},
and in the Yukawa model for $T/m_B = 1.4$\cite{KKN06b}, 
respectively.
There appears the three-peak structure in both models.
}
\label{fig:spectrum}
\end{figure}

The spectral functions of the quark and anti-quark are expressed as
$ \rho_\pm(\bm{p},\omega)\equiv -(1/\pi)$
${\rm Tr} [ {\rm Im} G^R ( \bm{p},\omega ) 
\gamma^0 \Lambda_\pm (\bm{p})]$, 
with the retarded Green function of the quark $G^R(\bm{p},\omega)$
and the projection operators 
$ \Lambda_\pm(\bm{q}) = (1\pm\gamma^0 \bm{\gamma}\cdot\bm{q}/|\bm{q}|)/2 $.
In the left panel of Fig.~\ref{fig:spectrum},
we show the quark spectral function $\rho_+(\bm{p},\omega)$
for $\varepsilon\equiv (T-T_c)/T_c = 0.1$, i.e., slightly above $T_c$.
One sees a three-peak structure in the spectral function. 
Our numerical calculation shows that 
the clear three-peak structure survives up to 
$\epsilon \simeq 0.2$\cite{KKN06}.

\section{Quark spectrum in Yukawa models}

We have seen that 
``elementary'' bosonic excitations with a small width 
significantly modify the quark spectrum at finite temperature.
To elucidate the mechanism of the
modification of the fermion spectra on the more general ground,
we employ the Yukawa model composed of 
a massless quark field $\psi$ and an elementary
 massive scalar boson $\phi$;
\begin{equation}
  {\cal L} = \bar{\psi} (i \partial \hspace{-0.53em} / \psi - g \phi) \psi
  + \frac12 \left( \partial_\mu \phi \partial^\mu \phi - m_B^2 \phi^2 \right).
\label{eq:lag}
\end{equation}
In this model, the quark self-energy in the imaginary time formalism
at the one-loop order is expressed as
\begin{eqnarray}
\tilde\Sigma ( \bm{p},i\omega_m )
= 
-g^2 T \sum_n \int \frac{ d^3\bm{k} }{ (2\pi)^3 }
{\cal G}_0 ( \bm{k},i\omega_n )
{\cal D}( \bm{p}-\bm{k} ,i\omega_m-i\omega_n )
\label{eq:tildeSigma}
\end{eqnarray}
where 
${\cal D} ( \bm{k},i\nu_n ) = [ (i\nu_n)^2 - \bm{k}^2 - m_B^2 ]^{-1}$
is the Matsubara Green function for the scalar boson.
Taking the Matsubara summation and the analytic continuation,
we obtain the self-energy in the real time.

In the right panel of Fig.~\ref{fig:spectrum},
we show $ \rho_\pm(\bm{p},\omega) $
at $T = 1.4m_B$ and $g=1$.
One sees a clear three-peak structure in the spectral function
at intermediate temperatures around
$ m_B < T < 2m_B $\cite{KKN06b},
as in the NJL model.

\section{Discussions}

\begin{figure}
\begin{center}
\includegraphics[width=0.45\textwidth]{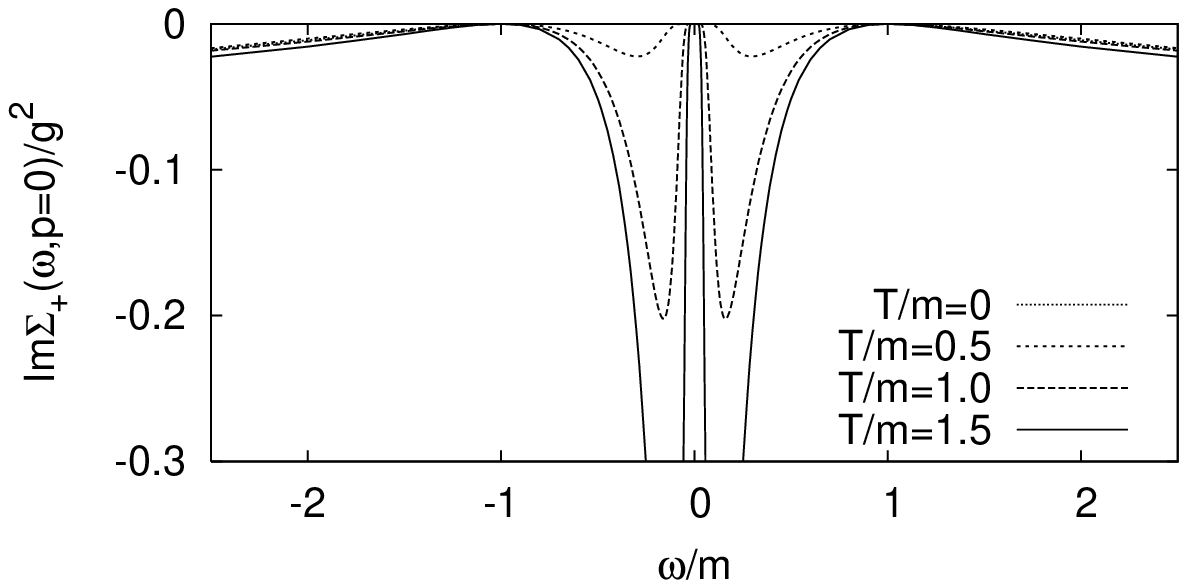}
\raisebox{10mm}{
\includegraphics[width=0.43\textwidth]{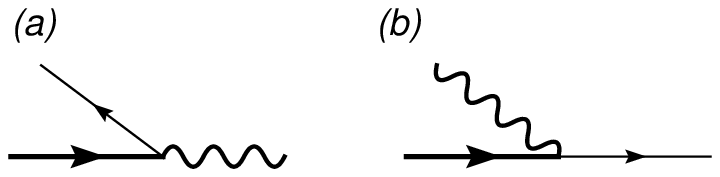}}
\end{center}
\vspace*{-10mm}
\caption[]{
Left panel:
The imaginary part of the quark self-energy
Im$\Sigma_+^R(\bm{0},\omega)$ in the Yukawa model
for several values of $T$. $m$ is the mass of the scalar boson.
Right figures:
Parts of the physical processes which 
Im$\Sigma_+^R$ describes.
The thick lines represent the quasi-quarks,
the thin lines the on-shell free quarks 
and the wavy lines the scalar boson.
}
\label{fig:sigma}
\end{figure}

In order to understand the mechanism of the novel structure
in $\rho_\pm( \bm{p},\omega )$,
we show the imaginary part of $\Sigma_+^R( \bm{p}=0,\omega )$ 
in the Yukawa model with several $T$ 
in the left panel in Fig.~\ref{fig:sigma}:
There are
two peaks in $|{\rm Im}\Sigma_+^R|$ 
with a positive and negative energy, which
grow rapidly as $T$ is raised.
It can be easily checked that 
these peaks 
in the positive and
negative energy regions physically correspond to
the Landau dampings shown
in the right panel in Fig.~\ref{fig:sigma}(a) and (b), respectively;
the wavy line represents the scalar boson.
We remark that the incident anti-quark line 
in Fig.~\ref{fig:sigma}(a) describes a thermally excited antiquark, 
which disappears after the collision with the scalar boson. 
The disappearance of the anti-quark means the
creation of a hole in the thermally-excited 
anti-quark distribution\cite{Weldon:1989ys}.
Figure~\ref{fig:sigma}(b) describes the decay process of a quasi-quark
which is a mixed state of quarks and antiquark-holes 
to an on-shell quark.
The two processes induce a respective quark-`antiquark hole' mixing
and hence two gaps are realized in $\rho_\pm$
at $\omega$ where Im$\Sigma^R_\pm$ has a peak 
corresponding to each process.

The mixing mechanism of the quarks can be understood 
in terms of the notion of {\em resonant scattering} 
as in the case of the (color-)superconductivity
\cite{JML97,KKN05}, although a crucial difference
arises owing to the different nature of the bosonic modes.
In the case of the superconductivity,
the precursory soft mode which induces the mixing is 
diffusion-mode like and has a strength around $\omega=0$\cite{KKKN02}.
The resonant scattering with the soft mode induces the mixing between
a particle and a hole, and gives rise to a gap-like
structure in the fermion spectrum around the Fermi 
energy\cite{KKKN04,KKN05}; 
correspondingly, the imaginary part $|{\rm Im}\Sigma^R|$ of the 
quark self-energy has a single peak around 
the Fermi energy\cite{KKN05}.
In the present cases,
the bosonic modes have a strength at finite energies
and hence the resonant scattering of the quarks 
bring the two peaks in $|{\rm Im}\Sigma^R|$ at finite energies
and induces a mixing between a quark (antiquark) and an
antiquark-hole (quark-hole).
Thus the two gaps in the quark spectrum are formed
in the positive- and negative-energy regions.

In summary, 
using the NJL model, we have shown that the quark spectrum 
in the QGP phase near the critical temperature of the chiral transition
can have a three-peak structure in the low-momentum region
owing to the effect of 
the hadronic soft modes of the chiral transition.
It has been shown that the three-peak structure in the fermion spectra 
also arises in the Yukawa model with an elementary massive boson.
The detailed analysis of this model
have elucidated the mechanism for the appearance
the three-peak structure, which is realized as a result of  
the mixing between a quark(anti-quark) and antiquark-hole (quark-hole)
induced by the {\em resonant scattering} of the quarks.

T.K. is supported by Grant-in-Aid
for Scientific Research by Monbu-Kagakusyo
(No. 17540250).
Y.N. is supported by the 21st Century COE progaram of 
Nagoya University and Grant-in-Aid
for Scientific Research by Monbu-Kagakusyo
(No. 18740140).

\end{document}